\documentclass{PoS}
\usepackage{amsfonts} 
\usepackage{amsmath} 
\usepackage{amssymb} 
\usepackage{latexsym} 
\usepackage{longtable} 
\usepackage{mathrsfs}
\usepackage{comment}

\newcommand{\mb}[1]{\mathbf{#1}}

\newcommand{\mr}[1]{\mathrm{#1}}
\newcommand{\bra}[1]{\langle#1|}
\newcommand{\ket}[1]{|#1\rangle}
\renewcommand{\mathcal}[1]{\mathscr{#1}}

\title{$D$ semileptonic form factors and $|V_{cs(d)}|$ from 2+1 flavor lattice QCD}

\ShortTitle{$D\rightarrow K(\pi)\ell\nu$ form factors and $|V_{cs(d)}|$ from lattice QCD}

\author{%
\speaker{Jon~A.~Bailey}$^{a,b}$,
A.~Bazavov$^c$,
A.X.~El-Khadra$^d$,
Steven~Gottlieb$^e$,
R.D.~Jain$^d$,
A.S.~Kronfeld$^a$,
R.S.~Van~de~Water$^c$,
and
R.~Zhou$^e$ \\
Email:  \email{jabailey@fnal.gov}\\ \\
\llap{$^a$}Theoretical Physics Department, Fermilab, Batavia, IL  60510, USA \\
\llap{$^b$}Department of Physics and Astronomy, Seoul National University, Seoul, 151-747, ROK\\
\llap{$^c$}Department of Physics, Brookhaven National Laboratory, Upton, NY
11973, USA\\
\llap{$^d$}Physics Department, University of Illinois, Urbana, IL  61801, 
USA \\
\llap{$^e$}Department of Physics, Indiana University, Bloomington, IN  47405, 
USA
}
\author{
Fermilab Lattice and MILC Collaborations\\}

\abstract{The measured partial widths of the semileptonic decays $D\rightarrow K\ell\nu$ and $D\to\pi\ell\nu$ can be combined with the form factors calculated on the lattice to extract the CKM matrix elements $|V_{cs}|$ and $|V_{cd}|$.  The lattice calculations can be checked by comparing the form factor shapes from the lattice and experiment.  We have generated a sizable data set by using heavy clover quarks with the Fermilab interpretation for charm and asqtad staggered light quarks on 2+1 flavor MILC ensembles with lattice spacings of approximately $0.12$, $0.09$, $0.06$, and $0.045$ fm.  Preliminary fits to staggered chiral perturbation theory suggest that we can reduce the uncertainties in the form factors at $q^2=0$ to below 5\%.}
          
\FullConference{XXIX International Symposium on Lattice Field Theory \\
                 July 10-16 2011\\
                 Squaw Valley, Lake Tahoe, California}

\begin{document}

\section{Introduction}
Performing a global fit under the assumption of a unitary CKM matrix~\cite{Charles:2011va} yields precise values for $|V_{cs}|$ and $|V_{cd}|$~\cite{Nakamura:2010zzi}.  If new physics in flavor introduces deviations from unitarity, however, values of the CKM matrix elements from direct determinations will in general differ from those predicted by the global fit.  Moreover, improving tests of second row and column unitarity requires improved direct determinations of $|V_{cs}|$~\cite{Nakamura:2010zzi}.

In the limit of massless leptons, the rates for the semileptonic decays $D\rightarrow K(\pi)\ell\nu$ become\footnote{For decays to the $\pi^0$, there is an additional isospin factor of $1/2$ on the right-hand side of Eq.~\eqref{eq:rate}.}
\begin{equation}
\frac{d\Gamma(D\to K(\pi)l\nu)}{dq^2}=\frac{G_F^2}{24\pi^3}|\mathbf{p}_{K(\pi)}|^3\,|V_{cs(d)}|^2|f_+^{D\to K(\pi)}(q^2)|^2,\label{eq:rate}
\end{equation}
where $q^2=(p_D-p_{K(\pi)})^2$, $\mathbf{p}_{K(\pi)}$ is the momentum of the $K(\pi)$ in the rest frame of the $D$, and $f_+^{D\rightarrow K(\pi)}(q^2)$ is defined in terms of the hadronic matrix element of the current $V_\mu=i\bar s\gamma_\mu c\ (i\bar d\gamma_\mu c)$:
\begin{equation}
\langle K(\pi)|V_\mu|D\rangle=f_+^{D\to K(\pi)}(q^2)\left(p_D+p_{K(\pi)}-\frac{m_D^2-m_{K(\pi)}^2}{q^2}q\right)_\mu + f_0^{D\to K(\pi)}(q^2)\,\frac{m_D^2-m_{K(\pi)}^2}{q^2}q_\mu.
\end{equation}
Given the normalization of the form factors $f_+^{D\rightarrow K(\pi)}(q^2)$ from lattice QCD, the CKM matrix elements $|V_{cs(d)}|$ can be extracted from experimental measurements of the branching fractions.

Agreement with the Standard Model values provides important validation of our methods, which we also use to calculate the form factors for $B\rightarrow \pi \ell\nu$~\cite{Bailey:2008wp} and $B\rightarrow K\ell^+\ell^-$~\cite{Zhou:2011be}.  Both decays are central in searches for new physics; the former allows direct extraction of $|V_{ub}|$, while the latter is loop-suppressed in the Standard Model.
New physics seems unlikely to enter the tree-level decays $D\to K(\pi)\ell\nu$ before the $B$ decays.  The Fermilab method applies to charm and bottom, so consistency between our $D$ form factors
and the normalizations implied by the global fit is direct evidence of our ability to precisely extract the $B$ form factors.

\section{\label{sec:meth}Method}
For the up, down, and strange quarks we use the asqtad-improved staggered action~\cite{ASQTAD}, for the charm quark we use the clover action with the Fermilab interpretation~\cite{HEAVY}, and for the gluons we use a one-loop Symanzik improved gauge action~\cite{GLUE}.  We set the scale with $f_\pi$, tune the masses of the light quarks using the experimental values of $m_\pi$ and $m_K$, and tune the hopping parameter of the charm quark with the spin-averaged kinetic mass of the $D_s$.

Table~\ref{tab:ens} summarizes our data set.  We vary the valence light-quark masses on each ensemble from near the tuned strange mass $m_s$ down to $\sim 0.1m_s$, and the lattice spacings from $\approx 0.12\ \mathrm{fm}$ to $\approx 0.045\ \mathrm{fm}$.  
To increase statistics and reduce autocorrelations, we average over four source times and randomize the source spatial locations.
\begin{table}[t]
\begin{center}
\begin{tabular}{cr@{$^3\times$}lrccr@{/}lr@{, }lc}
    \hline\hline
    $\approx a\ \mathrm{(fm)}$ & \multicolumn{2}{c}{$L^3\times n_t$} & $N_\mathrm{conf}$ & $n_\mathrm{src}$ & $n_\mathrm{snk}$ &  \multicolumn{2}{c}{$am_l/am_s$} &
    \multicolumn{2}{c}{$am_\mathrm{val}$} & $\kappa_c$ \\
    \hline 
    $0.12$ & 20&64 & 2052 & 4 & 4 & 0.02&0.05 & \{0.005&0.007, &
        0.1259 \\
           & 20&64 & 2259 & 4 & 4 & 0.01&0.05 & 0.01&0.02, &
        0.1254 \\
            & 20&64 & 2110 & 4 & 4 & 0.007&0.05 & 0.03&0.0415, &
        0.1254 \\
            & 24&64 & 2099 & 4 & 4 & 0.005&0.05 & 0.05&0.0349\} &
        0.1254 \\
    \hline 
    $0.09$ & 28&96 & 1996 & 4 & 4 & 0.0124&0.031 & \{0.0031&0.0047, &
        0.1277 \\
             & 28&96 & 1931 & 4 & 4 & 0.0062&0.031 & \multicolumn{2}{c}{0.0062,} &
        0.1276 \\
            & 32&96 & 984 & 4 & 4 & \hspace*{-6pt}0.00465&0.031 & 0.0093&0.0124, &
        0.1275 \\
            & 40&96 & 1015 & 4 & 4 & 0.0031&0.031 & 0.031&0.0261\} &
        0.1275 \\
            & 64&96 & 791  & 4 & 4 & \hspace*{-6pt}0.00155&0.031 & \multicolumn{2}{c}{} & 0.1275 \\
    \hline 
    $0.06$ & 48&144 & 593 & 4 & 4 & 0.0072&0.018 & \{0.0018&0.0025, &
        0.1295 \\
    & 48&144 & 673 & 8 & 4 & 0.0036&0.018 & \multicolumn{2}{c}{0.0036,} &
        0.1296 \\
        & 56&144 & 801 & 4 & 4 & 0.0025&0.018 & 0.0054&0.0072, & 0.1296 \\
      & 64&144 & 827 & 4 & 4 & 0.0018&0.018 & 0.0160&0.0188\} & 0.1296 \\
    \hline
    $0.045$ & 64&192 & 801 & 4 & 4 & 0.0028&0.014 & \{0.0018&0.0028, & 0.1310 \\
            & \multicolumn{2}{c}{} & \multicolumn{7}{r}{0.0040,
            0.0056, 0.0084,
            0.0160, 0.0130\}} \\ 
    \hline\hline
\end{tabular}
\end{center}
\caption{\label{tab:ens}Data on the 2+1 flavor asqtad staggered MILC ensembles for various valence masses and source times.  The columns are, respectively, the lattice spacing, lattice dimensions, number of configurations, number of source times, number of 3-point sink times, light/strange sea-quark masses, valence light-quark masses, and charm hopping parameter.  The analysis to date includes the full QCD points; we are considering generating data at additional source times.}
\end{table}

For calculations in the rest frame of the $D$ and in heavy-meson chiral perturbation theory, the hadronic matrix elements are conveniently parametrized by form factors $f_\bot$ and $f_{\|}$:
\begin{equation}
\bra{K(\pi)}V_\mu\ket{D}=\sqrt{2m_D}\left[v_\mu f_{\|}^{D\to K(\pi)}(q^2)+p_{\bot\mu}f_\bot^{D\to K(\pi)}(q^2)\right],\label{eq:ffdef}
\end{equation}
where $v=p_D/m_D$, and $p_\perp=p_{K(\pi)}-(p_{K(\pi)}\cdot v)v$.  $f_\bot$ and $f_{\|}$ can be extracted from correlator ratios.  We consider
\begin{equation}
{\overline R}_{3,\mu}^{D\to K(\pi)}(t,T;q^2)\equiv{\frac{1}{\phi_{K(\pi)\mu}}\frac{{\overline{C}}_{3,\mu}^{D\to K(\pi)}(t,T;\mb{p}_{K(\pi)})}{\sqrt {\overline{C}_2^{K(\pi)}(t;\mb{p}_{K(\pi)}){\overline{C}}_2^D(T-t)}}}\sqrt{\frac{2E_{K(\pi)}}{e^{-E_{K(\pi)}t}e^{-m_D(T-t)}}},\label{eq:ratio}
\end{equation}
where $\phi_{K(\pi)\mu}\equiv(1,\ \mb{p}_{K(\pi)})$, $E_{K(\pi)}=(m_D^2+m_{K(\pi)}^2-q^2)/(2m_D)$, and ${\overline C}_3,\ {\overline C}_2$ are averages of correlators constructed to eliminate oscillations from opposite-parity states~\cite{Bailey:2008wp}.  $T$ and $t$ are respectively the source-sink separation and current insertion time in the vector-current 3-points.

We use local operators for the $K(\pi)$ 2-points, smear the $D$ interpolators with a charmonium wavefunction, and construct the currents out of light staggered and heavy clover fields~\cite{Wingate:2002fh}.  For insertion times $t$ far from 3-point source and sink, the ratios plateau to the form factors:
\begin{equation}
{\overline R}_{3,0}^{D\to K(\pi)}\sim f_{\|}^{D\to K(\pi)}\quad\text{and}\quad{\overline R}_{3,i}^{D\to K(\pi)}\sim f_\bot^{D\to K(\pi)}\quad\text{for $1\ll t\ll T$}.
\end{equation}
The averages ${\overline C}_{3}$ require raw 3-points at successive source-sink separations $T$ and $T+1$; to minimize our errors as a function of momentum $\mathbf{p}_{K(\pi)}$, we generate the 3-point correlators at two physical separations on each ensemble and for each set of valence quark masses~\cite{Bailey:2009pz}.

We inject the 3-points and $K(\pi)$ 2-points with momenta $a\mathbf{p}_{K(\pi)}/(2\pi/L)=(0,0,0)$, $(1,0,0)$, $(1,1,0)$, $(1,1,1)$, and $(2,0,0)$ (and permutations and negatives of these components).  We average the correlators over equivalent momenta (up to axis interchange) and have checked that the wavefunction overlap factors of the $K(\pi)$ 2-points are independent of momentum.  We then replace $C_2^{K(\pi)}(\mathbf{p}_{K(\pi)})$ with the less noisy $C_2^{K(\pi)}(\mathbf{0})$ in the ratios of Eq.~(\ref{eq:ratio}) and use
\begin{equation}
{\overline R}_{3,\mu}^{\prime D\to K(\pi)}(t,T;q^2)\equiv\frac{1}{\phi_{K(\pi)\mu}}\frac{{\overline{C}}_{3,\mu}^{D\to K(\pi)}(t,T;\mb{p}_{K(\pi)})}{\sqrt{{\overline{C}_2^{K(\pi)}(t;\mb{0}}){\overline{C}}_2^D(T-t)}}\frac{E_{K(\pi)}}{e^{-E_{K(\pi)}t}}\sqrt{\frac{2e^{-m_{K(\pi)}t}}{e^{-m_D(T-t)}}}.\label{eq:new_ratio}
\end{equation}
We have checked that our data obeys the continuum dispersion relation, and we substitute this relation for $E_{K(\pi)}$ in the above ratio.

To extract the masses of the $K(\pi)$ and $D$ from the 2-point correlators, we fit to sums of exponentials with oscillating terms to account for contributions from opposite-parity states.  We use the masses to construct the ratios, and we propagate the errors {\it via} 500 bootstraps.  To extract the plateaus we fit the ratios; varying the fit function and time intervals does not significantly change the results.

The vector currents undergo renormalization.  We match to the continuum by writing the renormalization factors as products of the degenerate-mass vector-current normalization factors, which we compute nonperturbatively, and correction factors whose deviations from one are perturbatively calculable~\cite{ElKhadra:2001rv}:
\begin{equation}
\bra{K(\pi)}V_\mu\ket{D}=Z_{V_\mu}^{cs(d)}\bra{K(\pi)}V_\mu^\mathrm{lat}\ket{D},\quad Z_{V_\mu}^{cs(d)}=\rho_{V_\mu}^{cs(d)}\sqrt{Z_{V_4}^{cc}Z_{V_4}^{ss(dd)}}.
\end{equation}
We blind the analysis by 
introducing an offset in $\rho$.

\begin{figure}[tbph]
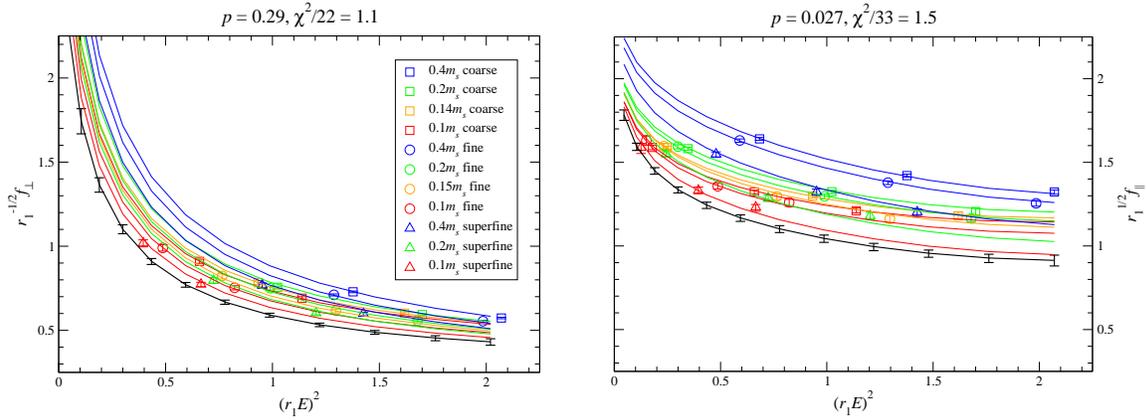

\begin{minipage}[c]{0.47\textwidth}
\begin{center}
\includegraphics[width=1.0\textwidth]{fperp_pion_11ens}
\end{center}
\end{minipage}
\hspace{0.05\textwidth}
\begin{minipage}[c]{0.47\textwidth}
\begin{center}
\includegraphics[width=1.0\textwidth]{fpar_pion_11ens}
\end{center}
\end{minipage}
\caption{\label{fig:fit_SChPT}Fits of $D\to\pi$ data to SU(3) S$\chi$PT; the legend applies to both plots.  Included are full QCD data from 11 ensembles at three lattice spacings and four light-quark masses.  The fit quality for $f_\bot^{D\to\pi}$ (left) is good, but the fit quality for $f_{\|}^{D\to\pi}$ is marginal (right).}
\end{figure}

After operator renormalization, the results for the form factors on all ensembles and for all combinations of valence masses and momenta are simultaneously fit to staggered chiral perturbation theory (S$\chi$PT)~\cite{Aubin:2007mc}.  Fits to SU(3) S$\chi$PT are shown in Fig.~\ref{fig:fit_SChPT}.  The data are for the $D\to \pi$ decay on a subset of the ensembles of Table~\ref{tab:ens}:  the four coarse ($a\approx 0.12$ fm) ensembles, four of the fine ($a\approx 0.09$ fm) ensembles (excluding the $0.05m_s$ ensemble), and three of the superfine ($a\approx 0.06$ fm) ensembles (excluding the $0.14m_s$ ensemble).  Data with momenta up through $\mathbf{p}_\pi=(1,1,0)$ are included in the fits.\footnote{Hereafter all momenta are given in units of $2\pi/(aL)$ unless otherwise specified.}  The fit function includes leading chiral logarithms (NLO loops) and analytic terms through NNLO.

The SU(3) S$\chi$PT fits for $f_\bot^{D\to\pi}$ and $f_\bot^{D\to K}$ are good, much better than SU(3) S$\chi$PT fits for $f_{\|}^{D\to\pi}$ and $f_{\|}^{D\to K}$.  As exemplified in Fig.~\ref{fig:fit_SChPT}, fits of $f_{\|}^{D\to\pi}$ data to SU(3) S$\chi$PT are marginal in quality.
We understand this behavior in terms of the features of the $\chi$PT description of the energy dependence of the form factors.

The pole from the resonance dominates the energy dependence of $f_\bot$, but the $\chi$PT expressions for $f_{\|}$ contain no pole, and the fits suffer.  The fit quality for $f_{\|}$ may also reflect the limitations of $\chi$PT.  For energies comparable to the chiral symmetry breaking scale, $E_{K(\pi)}\sim\Lambda_\mr{\chi SB}$, we expect the $\chi$PT description of the energy dependence to break down.  The largest-momentum data in fits to date have $\mb{p}_{K(\pi)}=(1,1,0)$, which corresponds to $E_\pi\sim\Lambda_\mathrm{\chi SB}$; on the coarse ensembles, $\mb{p}_\pi=(1,1,1)$ corresponds to energies $E_\pi\gtrsim \Lambda_\mr{\chi SB}$.

SU(3) S$\chi$PT fits for $f_{\|}^{D\to K}$ support the idea that the $\chi$PT description of the energy dependence of $f_{\|}$ is breaking down.  
For the same momenta, the energies $E_K$ are larger than the energies $E_\pi$, the $\chi$PT expansion breaks down for smaller momenta, and fits for $f_{\|}^{D\to K}$ should be worse than fits for $f_{\|}^{D\to\pi}$.
In fact, SU(3) S$\chi$PT fits for $f_{\|}^{D\to K}$ have unacceptably small $p$-values ($p<10^{-3}$).

To better parametrize the $f_{\|}$ data, we are investigating alternative fit functions, including SU(2) S$\chi$PT.  An improved treatment of the energy dependence of $f_{\|}$ would be especially desirable for adding data at higher momenta.  (Cf. Sec.~\ref{sec:err}.)  At the same time, $f_\bot$ dominates the desired form factor $f_+$, and SU(3) S$\chi$PT fits for $f_\bot^{D\to\pi}$ and $f_\bot^{D\to K}$ are good.  Any model dependence in $f_+^{D\to\pi}$ introduced by the $\chi$PT description of the energy dependence of $f_{\|}^{D\to\pi}$ is probably small.  We therefore proceed to compare the shape of the form factor $f_+^{D\to\pi}$ obtained from the fits shown in Fig.~\ref{fig:fit_SChPT} to the shape measured by CLEO.

\section{Comparison of lattice QCD and experiment}
If lattice calculations are to be combined with experimental results, the lattice and experimental results must be consistent.  Experiments measure shapes of form factors but cannot fix their normalizations without the CKM matrix elements~(Eq.~(\ref{eq:rate})).  For $D\to K(\pi)\ell\nu$, experiments and lattice calculations access the same $q^2$ region, and comparisons of the lattice and experimental form factor shapes provide stringent tests of lattice QCD.

\begin{figure}[tbph]
\begin{minipage}[c]{0.44\textwidth}
\begin{center}
\includegraphics[width=1.\textwidth]{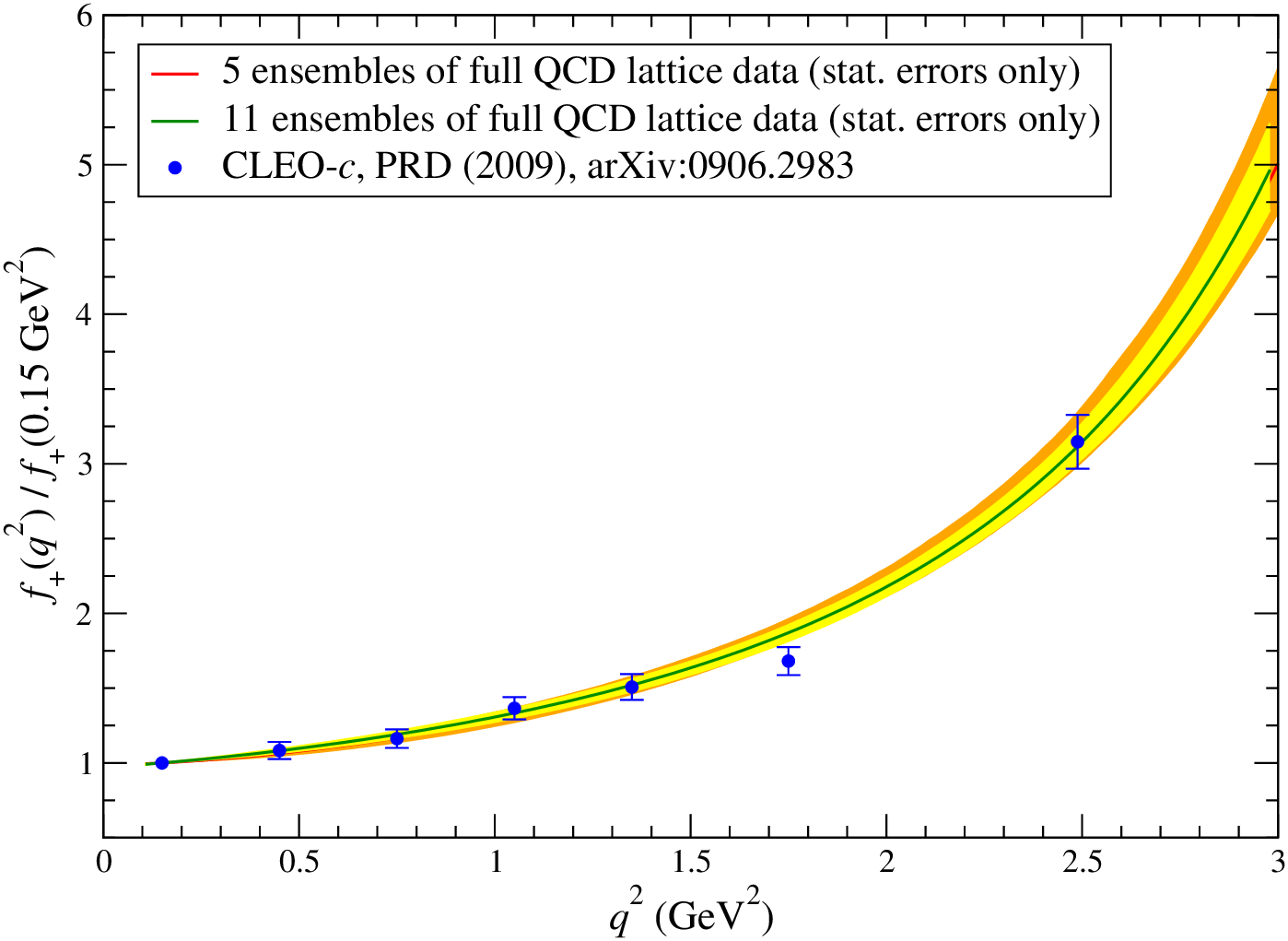}
\caption{\label{fig:overlay}Overlay of the ratio $f_+^{D\to\pi}(q^2)/f_+^{D\to\pi}({\tilde q}^2)$ from the lattice (curves and error bands) and CLEO (blue points)~\cite{Besson:2009uv}.  The larger lattice errors are from 5 ensembles, and the smaller are from 11 ensembles.  The form factor shapes from CLEO and the lattice agree; the errors scale as expected.}
\end{center}
\end{minipage}
\hspace{0.05\textwidth}
\begin{minipage}[c]{0.47\textwidth}
\begin{center}
\includegraphics[width=1.\textwidth]{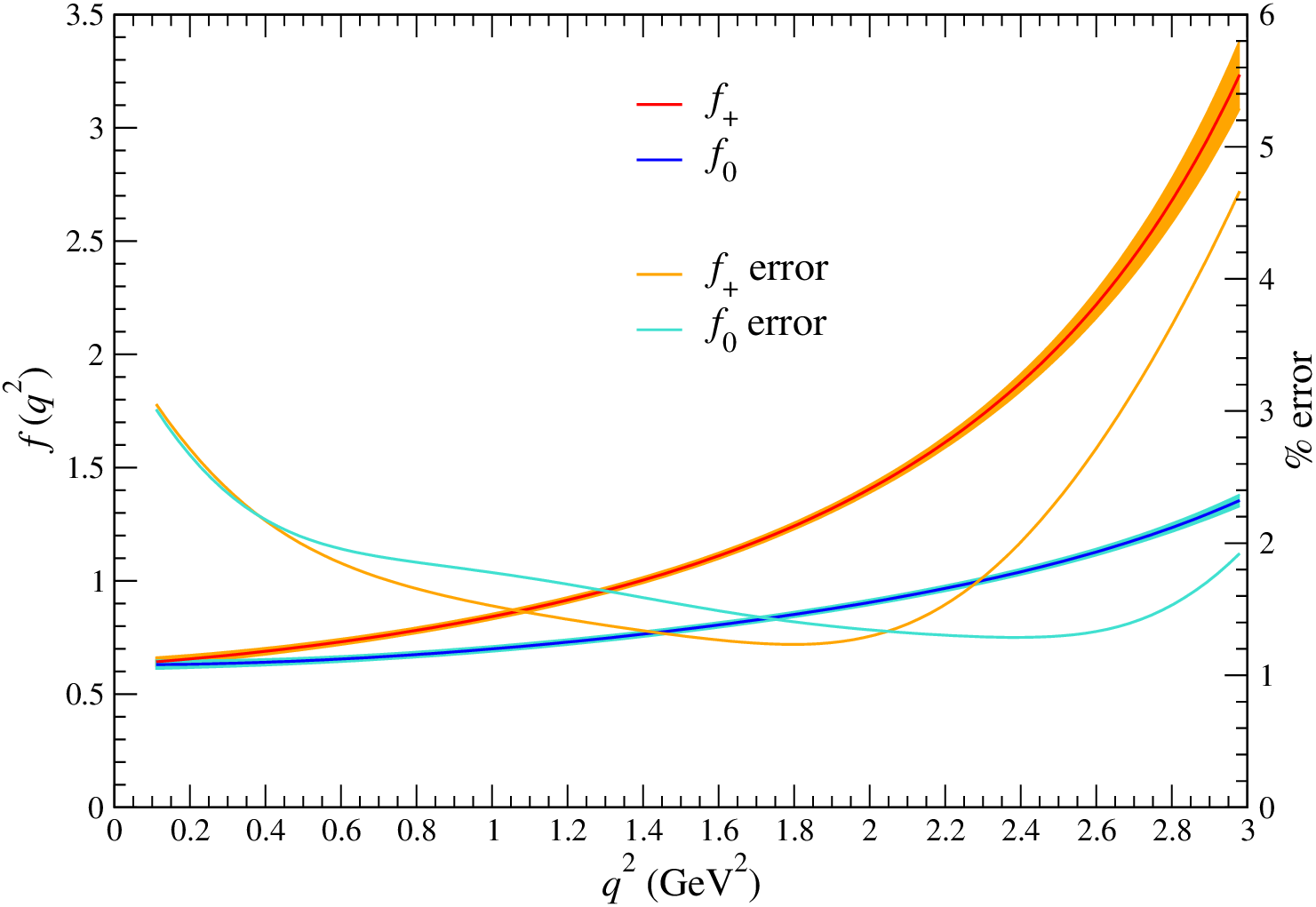}
\caption{\label{fig:ff_err}The form factors $f_+^{D\to\pi}(q^2)$ and $f_0^{D\to\pi}(q^2)$ and their statistical errors.  The curves are from S$\chi$PT, and the errors are from the Hessian.  The percent errors are plotted against the right-hand axis.  The larger errors at small $q^2$ probably reflect excluded data at corresponding momenta on the finer, more chiral ensembles.}
\vspace{1 mm}
\end{center}
\end{minipage}
\end{figure}

In Fig.~\ref{fig:overlay} we overlay our calculated $f_+^{D\to\pi}$, normalized to the point ${\tilde q}^2=0.15\ \mathrm{GeV}^2$, with the same ratio from CLEO~\cite{Besson:2009uv}.  The orange (dark grey) error band is the statistical error obtained by including data from two coarse and three fine ensembles~\cite{Bailey:2010vz}.  The yellow (light gray) error band is the statistical error obtained by including data from the 11 ensembles of Fig.~\ref{fig:fit_SChPT}.  The curves are from SU(3) S$\chi$PT, and the errors are from 500 bootstraps.  The errors scale as naively expected.  The form factor shapes from CLEO and the lattice calculation agree well.

\section{\label{sec:err}Projected errors}
To conservatively estimate our errors, we begin with the error budget of our $B\to\pi\ell\nu$ calculation~\cite{Bailey:2008wp}.  At $q^2=0$ naive scaling to the full QCD data set of Table~\ref{tab:ens} gives a statistical error of 4.2\% and an error from the degenerate-mass vector-current normalization factors of 0.6\%.  Updating the heavy-quark and $\rho$-factor power counting estimates to account for the ultrafine data gives errors of 2.5\% and 0.5\%, respectively.  For the remaining errors we adopt our previous estimates~\cite{Bailey:2009pz}.  This leads to a total error of 6.1\% for $f_+^{D\to K(\pi)}(0)$.
However, this estimate may be overly conservative.  In Fig.~\ref{fig:ff_err} we plot the form factors (with the $\rho$-factors set to one) $f_+^{D\to\pi}(q^2)$ and $f_0^{D\to\pi}(q^2)$ and their statistical errors as functions of $q^2$.  The data are from the 11 ensembles of Fig.~\ref{fig:fit_SChPT}.

The errors at $q^2_\mr{max}$ are much smaller for $f_0$ than for $f_+$ because $f_+$ ($f_0$) is dominated by $f_\bot$ ($f_{\|}$), and we have data for $f_\bot$ ($f_{\|}$) for $|\mb{p}_\pi|\ge 0.33\ \mr{GeV}\leftrightarrow q^2\le 2.0\ \mr{GeV^2}$ ($|\mb{p}_\pi|\ge 0\leftrightarrow q^2\le q^2_\mr{max}$).  As $q^2$ decreases, the errors reflect the addition of data and the hyperbolic behavior of the form factors; the errors in $f_+$ ($f_0$) grow approximately linearly in the region $1.6\ \mr{GeV^2}\gtrsim q^2\gtrsim 0.8\ \mr{GeV^2}$ ($2.0\ \mr{GeV^2}\gtrsim q^2\gtrsim 0.6\ \mr{GeV^2}$).  The largest momentum of data in the fit is $\mb{p}_\pi=(1,1,0)$, which corresponds to $q^2\in[0.43,1.58]\ \mr{GeV^2}$ on the ensembles with $m_l\le 0.2m_s$, and to $q^2\in[0.69,1.58]\ \mr{GeV^2}$ on the superfine ensembles.  Without data points below $q^2=0.43\ \mr{GeV^2}$ on the finer, more chiral ensembles, the errors increase rapidly as $q^2$ decreases in the region $0.3\ \mr{GeV^2}>q^2>0$.

Extrapolating the curve for the statistical error in $f_+$ to $q^2=0$, the error grows to about 3.6\%, somewhat smaller than that expected from naive scaling to the entire full QCD data set.  Including data at smaller $q^2$ would allow interpolation to $q^2=0$ and might improve the errors significantly.  We can appreciate the potential of the additional data by linearly extrapolating the curve for the error in $f_+$ for $q^2\in [0.8,1.6]\ \mr{GeV^2}$ to $q^2=0$.  The resulting expected error is about 2.0\%.  Adding a statistical error of 3.6\% (2.0\%) to our systematics yields a total error of 5.7\% (4.8\%).  Corresponding error budgets are in Table~\ref{table:err}.  We conclude that including data at momenta greater than $\mb{p}_{K(\pi)}=(1,1,0)$ may improve the error in $f_+(0)$ to better than 5\%.  This prospect further motivates us to consider alternatives to SU(3) S$\chi$PT for describing the energy dependence of $f_{\|}$ at small $q^2$.  (Cf. the last paragraph of Sec.~\ref{sec:meth}.)
\begin{table}[tbp]
\centering
\begin{tabular}{c||c|cccccccccc|c|c}
\hline\hline
 & Stat. & $g_{\pi}$ & $r_1$ & $\hat m$ & $m_s$ & $\kappa_c$ & $p_\pi$ & $\mathrm{HQ}$ & $Z_V$ & $\rho$ & FV & $\mathrm{Sys.}$ & $\mathrm{Tot.}$ \\
\hline
(a) & $3.6$ & $2.9$ & $1.4$ & $0.3$ & $1.3$ & $0.2$ & $0.1$ & $2.5$ & $0.6$ & $0.5$ & $0.5$ & $4.4$ & $5.7$ \\
(b) & $2.0$ & $2.9$ & $1.4$ & $0.3$ & $1.3$ & $0.2$ & $0.1$ & $2.5$ & $0.6$ & $0.5$ & $0.5$ & $4.4$ & $4.8$ \\
\hline\hline
\end{tabular}
\caption{\label{table:err}Projected error budgets for the form factors at $q^2=0$, assuming we (a) exclude data at momenta greater than $\mathbf{p}_{K(\pi)}=(1,1,0)$ and (b) include data at momenta greater than $\mathbf{p}_{K(\pi)}=(1,1,0)$.  Errors are due to limited statistics and the truncation of S$\chi$PT; uncertainties in the $D^*D\pi$ coupling, scale, average up-down quark mass, strange quark mass, and charm hopping parameter; momentum-dependent discretization effects of the light quarks and gluons; heavy-quark discretization effects; uncertainties in the matching factors $Z_V$ and $\rho$; and finite volume effects.  The last two entries are the total systematics and total error.}
\end{table}

\medskip
Computations were carried out with resources
provided by the USQCD Collaboration, Argonne Leadership
Computing Facility, and National Energy Research
Scientific Computing Center, which are funded by the Office of
Science of the U.S. DOE; and with resources
provided by the National Institute for Computational Science,
Pittsburgh Supercomputer Center, San Diego Supercomputer 
Center,
and Texas Advanced Computing Center, which are funded
through the National Science Foundation's Teragrid/XSEDE
Program. This work was supported in part by the U.S.
DOE under grant Nos.
DE-FG02-91ER40661 (S.G.), No. DE-FG02-91ER40677 (R.D.J.,
A.X.K.).
This manuscript was co-authored by employees of
Brookhaven Science Associates, LLC, under Contract No.
DE-AC02-98CH10886 with the U.S. DOE. R.S.V.
acknowledges support from BNL via the Goldhaber Distinguished
Fellowship.
J.A.B. is supported by the Creative Research
Initiatives program (3348-20090015) of the NRF grant funded by the
Korean government (MEST).
Fermilab is operated by Fermi Research Alliance, LLC, under
Contract No. DE-AC02-07CH11359 with the U.S.
DOE.

\end{document}